\documentclass[
]{ceurart}
\usepackage{comment}
\usepackage{todonotes}
\usepackage{hyperref}
\usepackage{listings}
\begin{document}

\copyrightyear{2022}
\copyrightclause{--}
\conference{--}

\title{Mapping STI ecosystems via Open Data: overcoming the limitations of conflicting taxonomies. A case study for Climate Change Research in Denmark}



%
\author[1]{Nicandro Bovenzi}[%
orcid=0000-0002-7764-3974,
]
\address[1]{SIRIS Lab, Research Division of SIRIS Academic, 08003, Barcelona, Spain}

\author[1]{Nicolau Duran-Silva}[%
orcid=0000-0001-5170-4129,
]

\author[1]{Francesco Alessandro Massucci}[%
orcid=0000-0001-5405-8759,
email=francesco.massucci@sirisacademic.com,
]
\cormark[1]

\author[1]{Francesco Multari}[%
orcid=0000-0001-8509-3178,
]

\author[1]{C\'esar Parra-Rojas}[%
orcid=0000-0003-3625-9412,
]

\author[1]{Josep Pujol-Llatse}[%
]
\cortext[1]{Corresponding author.}

\begin{abstract}
Science, Technology and Innovation (STI) decision--makers often need to have a clear vision of {\em what} is researched and by {\em whom} to design effective policies. Such a vision is provided by effective and comprehensive mappings of the research activities carried out within their institutional boundaries. A major challenge to be faced in this context is the difficulty in accessing the relevant data and in combining information coming from different sources: indeed, traditionally, STI data has been confined within closed data sources and, when available, it is categorised with different taxonomies. Here, we present a proof--of--concept study of the use of Open Resources to map the research landscape on the Sustainable Development Goal (SDG) 13 -- Climate Action, for an entire country, Denmark, and we map it on the 25 ERC panels.
\end{abstract}



\maketitle

\section{Introduction}
\vspace{-.25cm}
To inform their decisions, policy-makers in the Science, Technology and Innovation (STI) sector typically need ``maps'', either at a territorial or at an institutional level, to understand what is researched and by whom. Generally, those maps need to provide information about the research and innovation topics and about the relevant actors linked to them, so that effective policy-actions could be proposed, by covering the right scientific domains and by being catered for the adequate users. These maps need to be comprehensive to extensively cover {\em i.} the whole STI value chain (from basic research up to industrial innovation), {\em ii.} the different scientific domains and  {\em iii.} all possible relevant actors. As such, these maps should rely on different data sources that could offer the broadest possible view of STI inputs and outputs. 
Some major challenges faced at a policy level arise because many of those data sources are not openly available (undermining therefore the participatory processes), they are not interoperable in terms of data classification schemes and institutional identification (therefore limiting transversal analyses) and they are hardly manageable by non--expert users \cite{Fuster2020}.


In this paper, we present a proof of concept of merging different open datasets and of analysing them with a common classification scheme specifically designed for the sake of our analyses. To do so, we gather data for the whole Danish\footnote{We choose Denmark as our case study because (i) it is a medium-size country and the size of its scientific production is such that one can practically retrieve all the documents from publications repositories, (ii) Danish R\&D ecosystem is internationally visible and competitive and (iii) Denmark is internationally acknowledged as being one of the leading countries in terms of climate action policies and efforts (1st in the world according to the 2022 Environmental Performance Index (EPI)).} STI ecosystem from 4 different data sources, namely:  
{\em i.} the CORDIS database\footnote{\url{https://cordis.europa.eu/}}, through the UNICS \cite{Gimenez2018UNiCSTO} platform, for H2020--funded R\&D projects, {\em ii.} the Kohesio linked--data portal\footnote{\url{https://kohesio.ec.europa.eu/}} for Regional R\&D projects funded by the EC Cohesion initiative, and the {\em iii.} OpenAlex \cite{openalex} and {\em iv.} OpenAIRE\cite{openaire} repositories for publications and other scientific outputs (these last being accessed through their respective APIs). After gathering these records, we use open knowledge--bases \cite{duran2019controlled} and fine tune openly available Deep Learning models \cite{Sun2019HowTF, cohan2020specter} to:
    {\bf {\em i.} Identify STI documents linked with the Sustainable Goal 13}, Climate Action \cite{duran2019controlled}, and
    {\bf {\em ii.} Categorise} Danish research on Climate Action {\bf within the 25 panels of the European Research Council} (ERC).

In this way, we aim at showcasing how research in emerging fields (such as the SDGs \cite{sdg2019sustainable}) can be gathered from open data sources and identified by means of modern, openly available AI models \cite{chesbrough2006open}. Finally, we demonstrate how gaps in taxonomic classifications across datasets may be filled by means of Deep Learning textual classifiers, by using the ERC panels as a paradigmatic example.

\section {Main Results}

We gathered, from a series of heterogeneous data sources, a dataset of scientific publications abstracts and R\&D projects descriptions for the entire STI ecosystem of Denmark for the 2014--2019 time period. We then tagged each single record by means of a controlled vocabulary for SDG 13 -- Climate Action \cite{duran2019controlled} (that is, we identified the vocabulary terms in each single text by applying a series of textual matching rules). This enabled us to identify, within the initial dataset, all textual records linked with SDG 13. 

The number of documents that we could identify in each data source as well as those we could link with SDG 13 are reported in Table~\ref{results_records}. About 2\% of Scientific publications in Denmark between 2014 and 2019, both from OpenAIRE and OpenAlex, are related to our SDG of interest. In contrast, European projects are more linked, in relative terms, to the issue of Climate Action: this comes as no surprise, given the orientation of EU policies towards the sustainability issues.

\begin{table}[htt]
\caption{Number of records with at least one author affiliation or beneficiary from Denmark (2014-2019) and relative volume mapped to SDG 13.}

\addtolength{\tabcolsep}{1pt}
\begin{tabular}{|l|r|r|}
\hline
\textbf{Data source} & \textbf{Total Records in DK} & \textbf{Records related to SDG 13} \\
\hline
OpenAlex & 191,399 & 3,821 (2\%) \\
OpenAIRE & 235,906 & 5,273 (2.2\%) \\
CORDIS & 2,196 & 320 (14.6\%) \\
Kohesio & 294 & 14 (4.8\%) \\
\hline
\end{tabular}
\addtolength{\tabcolsep}{-1pt}
\label{results_records}
\end{table}

\begin{figure}[h]
  \centering
  \includegraphics[width=\linewidth]{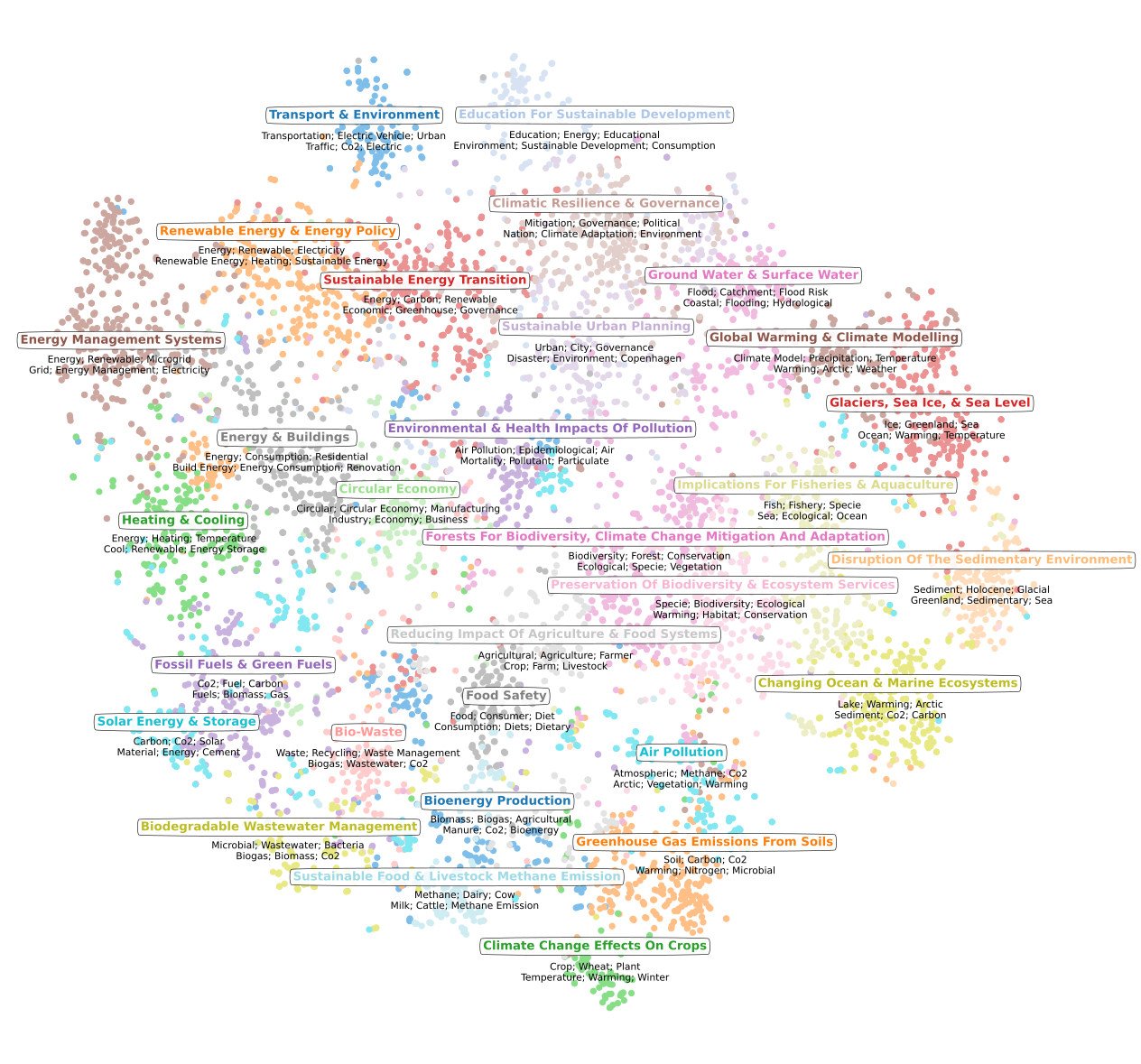}

\caption{t-SNE visualisation of the 30 topics extracted from the SDG 13 corpus.} 
\label{confetti}
\end{figure}

We finally applied Topic Modelling \cite{griffiths2004finding, topicbert1} and we classified per ERC panels the SDG--related corpus, in order to obtain both a series of topics characterising Climate Action--related research and to gain a ``disciplinary'' view of such research. In Fig.~\ref{confetti}, we show a t-SNE visualisation of the automatically extracted topics \cite{van2008visualizing} from textual data in publications (from OpenAlex and OpenAIRE) and projects (from CORDIS and Kohesio) concerning SDG 13. 

In Fig.~\ref{confetti}, each dot is a single document: as one can observe, we were able to extract a series of thematically different topics from the SDG 13 corpus, each dealing with a different aspect of climate action. At the centre of the figure, one finds topics related with the environment, while going anti--clockwise from the top, it is possible to encounter topics related with energy, traditional and alternative fuels, emissions and pollution, impact on the biosphere and finally education and policy issues. 

\begin{figure}[h]
\begin{center}
  \includegraphics[width=0.9\linewidth]{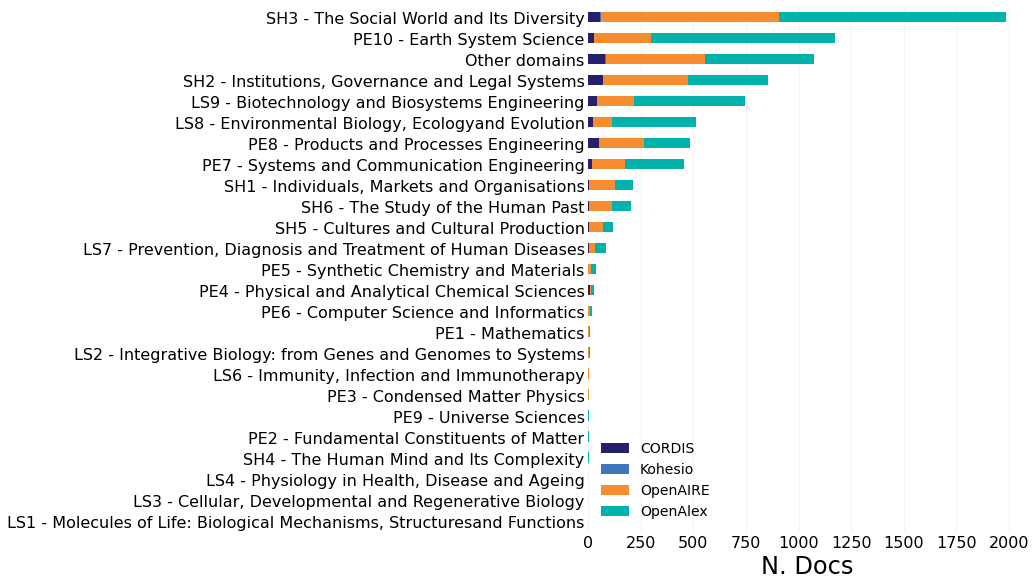}
\end{center} 
\caption{Number of documents per source and by ERC Panel.}
\label{number_erc_source}
\end{figure}

As a final pilot exercise, we proceeded to classify the documents linked to SDG 13 per ERC panels. To do so, we trained a Deep Learning textual classifier by fine-tuning \cite{devlin2018bert, Sun2019HowTF} the BERT-based SPECTER model \cite{cohan2020specter} on a weakly supervised dataset, and we applied it to our Danish SDG 13 corpus. This effort allowed us to obtain a disciplinary classification of the records which is consistent across data sources and which may enable, in turn, a comparison of the Danish STI ecosystem with other geographical perimeters of interest. In Fig.~\ref{number_erc_source}, we present the distribution of documents by source and ERC panel. Perhaps surprisingly, the majority of the STI documents analysed were linked to Social Science issues, followed by Earth Sciences. Also, interestingly (and underscoring the importance of cross--platform analyses such as this one), one can see that the various data sources have a different coverage of the panels.

\section{Conclusions}

In this paper, we presented a proof--of--concept study of the use of Open Resources to map the research landscape on SDG 13 (Climate Action), for an entire country, Denmark. This type of mapping exercise is extremely useful for STI decision--makers, who, to design effective policies within their respective sphere of influence, need to have a clear vision of {\em what} is researched and by {\em whom}.

Here, we carried out a study of this sort by relying on Open Data for Research Projects (gathered from CORDIS and the Kohesio platform) and Scientific publications (collected from OpenAIRE and OpenAlex), by using an open vocabulary for mapping STI records on SDG 13 and by using openly available Deep Learning models to classify the corpus in accordance with the 25 ERC panels. The results we obtain are fairly encouraging: the coverage of the data analysed is extensive, both in absolute terms\footnote{We made an overall comparison with results produced by Scopus: the number of records was much lower than what found both in OpenAIRE and OpenAlex, for the same country and time period} and in terms of scientific disciplines and actors. Interestingly, the data sources analysed offer a complementary view of the research domains, and allow one, when used in combination, to obtain a wide and precise overview of the local STI ecosystem.

\section*{Acknowledgments}
This work was partly funded by the European Commission H2020 Programme via the INODE project, under grant agreement No 863410.

\bibliography{bibliography}
\end{document}